\documentclass[prd,twocolumn,showpacs,amssymb]{revtex4}

\usepackage{graphicx}
\usepackage{dcolumn}
\usepackage{bm}

\begin{document}

\title{Chromomagnetic instability in dense quark matter}

\author{Mei Huang}

\author{Igor A. Shovkovy}
\altaffiliation[On leave from: ]{%
       Bogolyubov Institute for Theoretical Physics,
       03143, Kiev, Ukraine}%

\affiliation{%
       Institut f\"{u}r Theoretische Physik,
       J.W. Goethe-Universit\"{a}t,
       D-60054 Frankurt/Main, Germany}%

\date{\today}

\begin{abstract}
The results for the Debye and Meissner screening masses of the gluons 
and the photon in the case of neutral and $\beta$-equilibrated dense
two-flavor quark matter are presented. In the limits of the normal 
phase and the ideal two-flavor color superconducting phase, the
screening masses coincide with the known results. Most interestingly, 
we find that the Meissner screening masses squared can be {\em negative}, 
indicating a plasma-type chromomagnetic instability in dense quark matter. 
\end{abstract}



\maketitle

\section{Introduction}

It is known that sufficiently cold and dense baryonic matter is a 
color superconductor. It is possible that such a state of matter 
may naturally exist in the Universe inside central regions of 
compact stars. For this reason, the topic of color superconductivity 
stirred a lot of interest in recent years \cite{cs,cfl,weak,weak-cfl}. 
(For reviews on color superconductivity see, for example, 
Ref.~\cite{reviews}.)

Matter in the bulk of stars is electrically neutral and 
$\beta$ equilibrated. 
This means that the chemical potentials of different quarks satisfy 
some nontrivial relations. Such relations affect the pairing dynamics 
between quarks, and the ground state of matter is modified. It was 
argued in Refs.~\cite{absence2sc,neutral_steiner}, for example, 
that a mixture of the two-flavor color superconducting (2SC) phase and 
unpaired strange quarks is less favorable than the color-flavor-locked 
(CFL) phase when the charge neutrality is enforced.

When the constituent mass of the strange quark is large, neutral 
two-flavor quark matter in $\beta$ equilibrium can have a ground 
state called the gapless color superconductor (g2SC) \cite{SH}. 
While the symmetry of the g2SC ground state is the same as 
that of the conventional two-flavor color superconductor (2SC), 
the spectrum of the fermionic quasiparticles is different. 

The existence of stable gapless color superconducting phases was
confirmed in Refs.~\cite{GLW,var-appr,RD}, and generalized to finite 
temperatures in Refs.~\cite{HS,LZ,Iida}. It was also shown that a
gapless color-flavor-locked (gCFL) phase can appear in neutral strange 
quark matter \cite{gCFL,RSR}. A nonrelativistic analogue of gapless 
superconducting phases could appear in trapped gases of cold 
fermionic atoms \cite{WilLiu,Deb,LWZ,FGLW}. (Note that a mixed 
phase can be an alternative ground state in atomic \cite{Bed} 
as well as quark \cite{neutral_buballa,SHH,RR} systems provided 
the surface tension is small.)

In this paper, we report on the gluon screening properties in
neutral dense two-flavor quark matter. Our results cover the 
gapped as well as the gapless 2SC phases. In the case of 
the ideal 2SC phase (i.e., without a mismatch between the Fermi 
momenta of different quarks), our results reproduce the findings
of Refs.~\cite{R-meissner,SWR}. The most important result of this 
paper is the finding of a chromomagnetic type plasma instability 
in neutral dense quark matter. We show that there exist five 
unstable gluon modes 
in the g2SC phase, and four of them persist even in the gapped 
phase when the mismatch is larger than a certain critical value. 
We argue that this instability may lead to a gluon condensation 
in dense quark matter. Specific details of the condensate still
remain to be clarified.

\section{Gluon self-energy}
\label{gpt}

In the ground state of a two-flavor color superconductor, the gauge 
symmetry group ${\rm SU}(3)_c \times {\rm U}(1)_{\rm em}$ is broken 
down to ${\rm SU}(2)_c \times \tilde{\rm U}(1)_{\rm em}$ by the 
Anderson-Higgs mechanism. Normally, this should lead to a generation 
of masses for the 5 gluons that correspond to the broken generators.
This would be a standard description of the Meissner effect. As we shall
see below, this is not always the case in dense quark matter.

In order to study the gluon screening properties in the system at 
hand, we follow an approach similar to that in Refs.~\cite{R-meissner,SWR}.
The polarization tensor in momentum space has the following general 
structure:
\begin{equation} 
\Pi^{\mu \nu}_{AB} (P) = 
-\frac{i}{2} \int\frac{d^{4}K}{(2\pi)^{4}}
\mbox{Tr}_{\rm D} \left[ \hat{\Gamma}^\mu_A
{\cal S} (K) \, \hat{\Gamma}^\nu_B {\cal S}(K-P) \right].
\label{PiPscfNG}
\end{equation}
The trace here runs over the Nambu-Gorkov, flavor, color and Dirac
indices. The 4-momenta are denoted by capital letters, e.g., $P=
(p_0,\vec{p})$. The explicit form of vertices $\hat{\Gamma}^\mu_A$
is 
\begin{eqnarray}
\label{vertex}
\hat{\Gamma}_A^\mu
\equiv \left\{\begin{array}{lll} 
{\rm diag}(g\,\gamma^\mu T_A,-g\,\gamma^\mu T_A^T) & 
\mbox{for} &  A=1,\ldots,8 \, , \\ \\
{\rm diag}(e\, \gamma^\mu Q,-e\,\gamma^\mu Q) & \mbox{for} &  A=9 \, ,
\end{array} \right.    
\end{eqnarray}
where $T_A$ and $Q$ are the generators of the color and the 
electromagnetic gauge transformations (with the $A=9$ gauge 
boson corresponding to the photon). Note that the strong and 
electromagnetic coupling constants, $g$ and $e$, are 
included in the definition of the vertices. The inverse 
of the quark propagator is defined as
\begin{equation}
\left[S(P)\right]^{-1} = \left(\begin{array}{cc}
\left[G_0^{+}(P)\right]^{-1} & \Delta^- \\
\Delta^+ & \left[G_0^{-}(P)\right]^{-1}
\end{array}\right).
\end{equation}
with the off-diagonal elements
$\Delta^{-} = -i \epsilon^{b}\varepsilon\gamma^5 \Delta$ and
$\Delta^{+} \equiv \gamma^0 \left(\Delta^{-}\right)^{\dagger} \gamma^0
= -i \epsilon^{b}\varepsilon\gamma^5\Delta^{*}$.
Here $\Delta$ is the diquark gap parameter, while 
$\left(\epsilon^{b}\right)_{\alpha\beta} \equiv \epsilon^{b\alpha\beta}$
and $\left(\varepsilon\right)^{ij}\equiv  \varepsilon^{ij}$ are 
antisymmetric tensors in the color and flavor spaces, respectively. 
Without losing generality, we assume that the quarks are massless 
in dense quark matter \cite{neutral_steiner,neutral_huang}. Then 
the free quark propagators 
$G_0^{\pm}(P)$ read
\begin{equation}
\left[G_0^{\pm}\right]^{-1} =
 \gamma^0 (p_0-p \pm \hat{\mu})\Lambda^{+}_{p}
+\gamma^0 (p_0+p \pm \hat{\mu})\Lambda^{-}_{p},
\end{equation}
where $\Lambda^{\pm}_{p}\equiv \frac{1}{2}
\left[1\pm \gamma^0 (\vec{\gamma} \cdot \vec{p})/p\right]$ 
are the positive and negative energy projectors, and 
$\hat{\mu}$ is the matrix of chemical potentials in the color and 
flavor space. In $\beta$ equilibrium, the nontrivial elements of 
matrix $\hat{\mu}$ read \cite{SH,HS}
\begin{eqnarray}
\mu_{u r} =\mu_{u g} =\mu -\frac{2}{3}\mu_{e} +\frac{1}{3}\mu_{8}, \\
\mu_{d r} =\mu_{d g} =\mu +\frac{1}{3}\mu_{e} +\frac{1}{3}\mu_{8}, \\
\mu_{u b} =\mu -\frac{2}{3}\mu_{e} -\frac{2}{3}\mu_{8}, \\
\mu_{d b} =\mu +\frac{1}{3}\mu_{e} -\frac{2}{3}\mu_{8},
\end{eqnarray}
with $\mu$, $\mu_{e}$, and $\mu_{8}$ being the chemical potentials of
the quark number, the electrical charge and the color charge, respectively. 

\section{Screening masses of gauge bosons}
\label{screening}

The Debye masses $m_{D,A}^2$ and the Meissner masses $m_{M,A}^2$ of 
gauge bosons are defined in terms of the eigenvalues of the 
polarization tensor \cite{R-meissner,SWR}. In the basis in which 
$\Pi_{AB}^{\mu\nu}(0,p)$ is diagonal, they become
\begin{eqnarray}
m_{D,A}^2 &\equiv & - \lim_{p\to 0} \tilde{\Pi}_{AA}^{00}(0,p), 
\label{def-Debye}\\
m_{M,A}^2 &\equiv & - \frac{1}{2}\lim_{p\to 0} 
\left(g_{ij}+\frac{p_i p_j}{p^2}\right) \tilde{\Pi}_{AA}^{ji}(0,p) .
\label{def-Meissner}
\end{eqnarray}
Below, we give only the final results for these quantities. The 
details of the calculation will be presented elsewhere \cite{pi-long}.

As was shown in Refs.~\cite{SH,HS}, the ground state of neutral 
dense quark matter is determined by the strength of the diquark 
coupling constant. At weak diquark coupling, the Cooper pairing is in
conflict with the constraint of charge neutrality. As a result, the 
ground state corresponds to the normal phase. At strong diquark coupling, 
on the other hand, the ground state is in the gapped 2SC phase, and 
the neutrality plays little effect. In the regime of an intermediate 
strength of the coupling, the ground state is given by the gapless 
2SC phase \cite{SH,HS}. 

In order to describe the most general situation, we parametrize all 
the results below with the dimensionless ratio $\Delta/\delta\mu$, 
where $\delta\mu\equiv \mu_e/2$ is the mismatch between the Fermi 
momenta of paired quarks which is determined by the neutrality and 
the $\beta$-equilibrium conditions. The value $\Delta/\delta\mu=0$ 
corresponds to the normal phase, while $\Delta/\delta\mu<1$ and 
$\Delta/\delta\mu>1$ correspond to the gapless and the gapped 2SC 
phases, respectively \cite{SH,HS}.

\begin{figure}
\noindent
\includegraphics[width=0.49\textwidth]{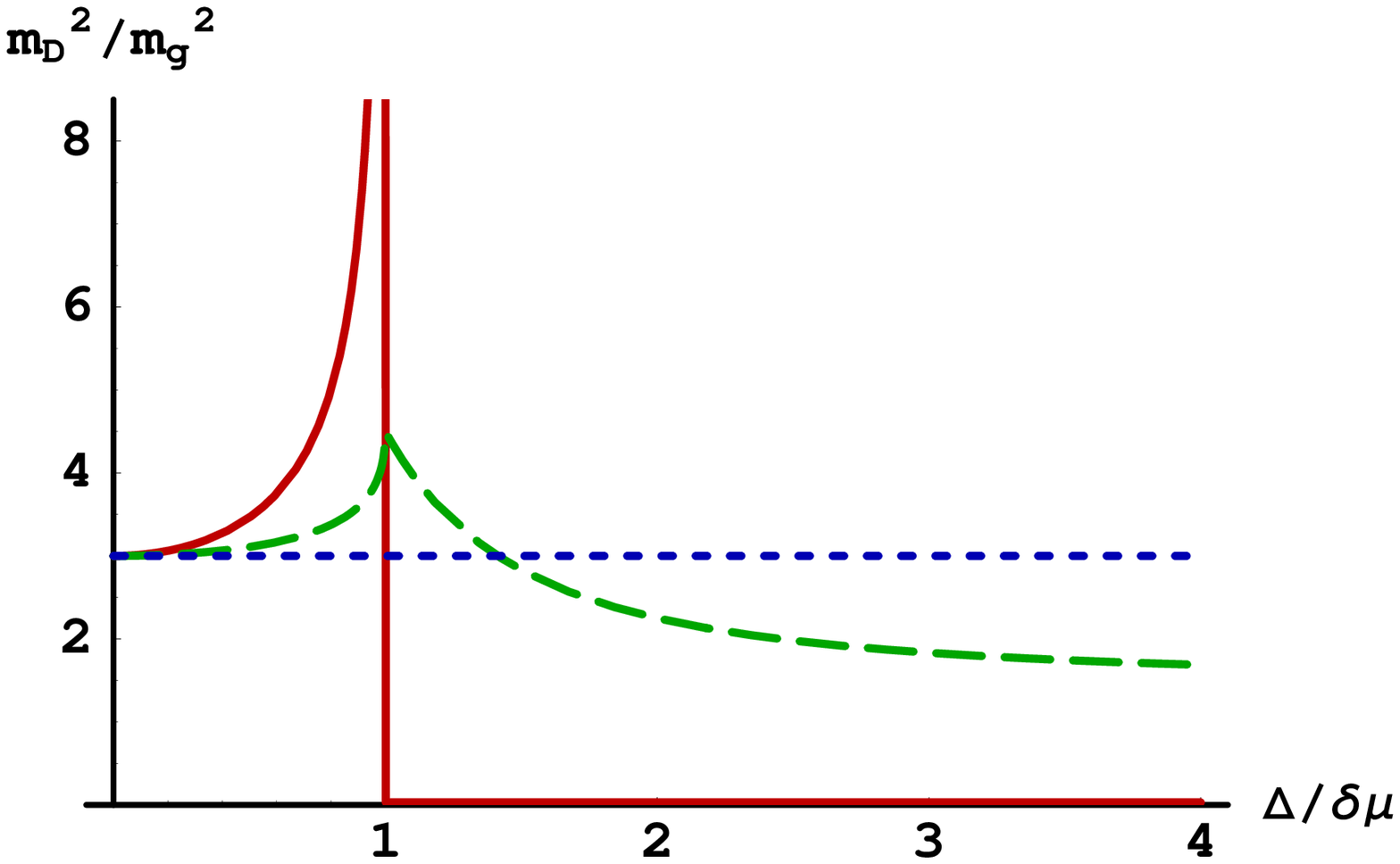}\\
\noindent
\includegraphics[width=0.49\textwidth]{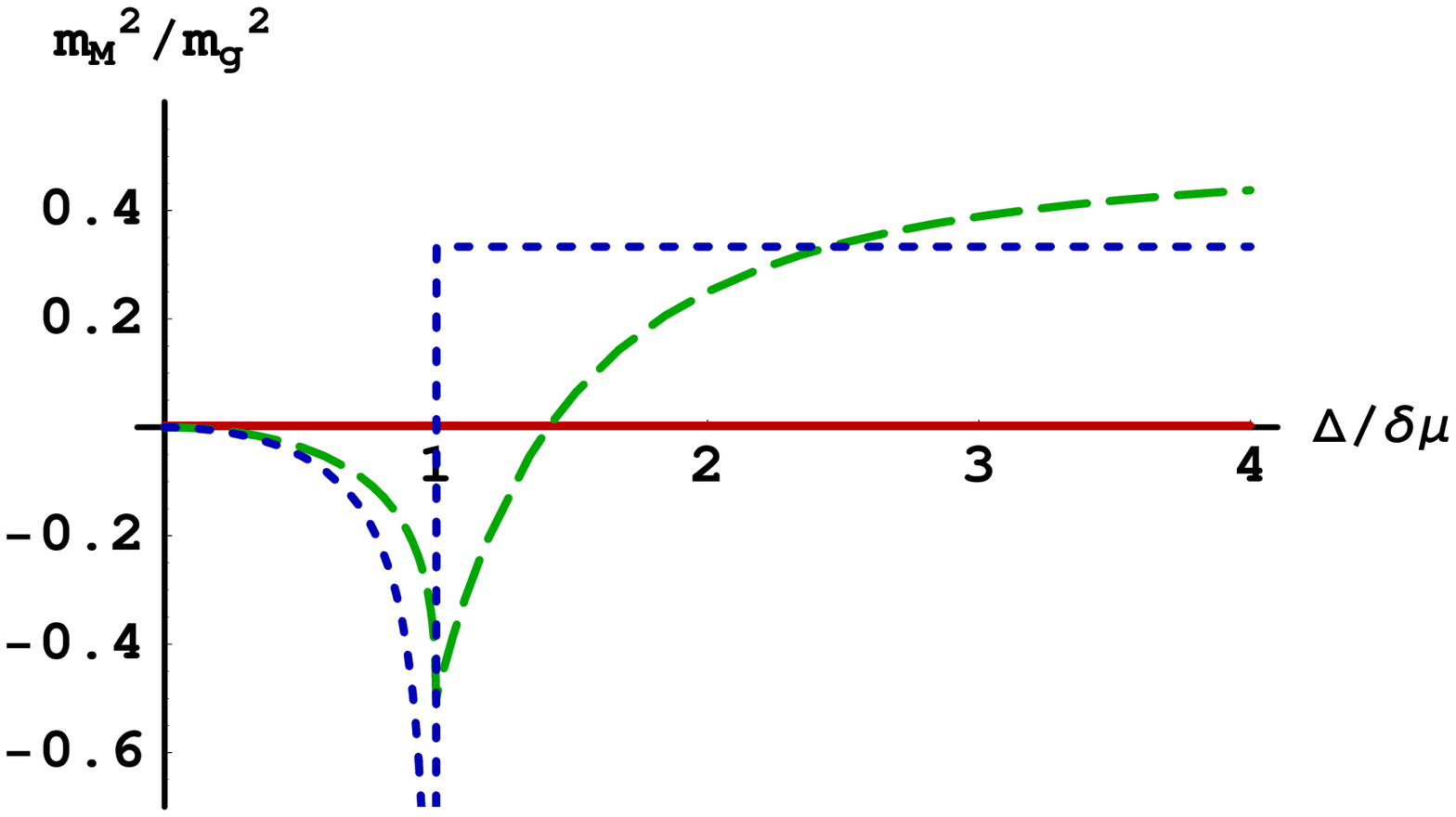}
\caption{
Squared values of the gluon Debye (upper panel) and Meissner 
(lower panel) screening masses, devided by $m_g^2
=4\alpha_s\bar\mu^2/3\pi$, as functions of the dimensionless 
parameter $\Delta/\delta\mu$. The red solid  line denotes the 
results for the gluons with $A=1,2,3$, the green long-dashed
line denotes the results for the gluons with $A=4,5,6,7$, and 
the blue short-dashed line denotes the results for the gluon 
with $A=\tilde{8}$.}
\label{fig-Masses}
\end{figure}

{\em Screening masses of the gluons with $A=1,2,3$.}
The general expression for the polarization tensor 
$\Pi_{AB}^{\mu\nu}(0,p)$ with $A,B=1,2,3$ is diagonal.
Then, by making use of the definition in Eq.~(\ref{def-Debye}),
we arrive at the following result for the threefold degenerate 
Debye mass:
\begin{eqnarray}
m_{D,1}^2 &=& \frac{2\alpha_s}{\pi}
\frac{\left[(\mu^{-})^2+(\mu^{+})^2\right]
\delta\mu}{\sqrt{(\delta\mu)^2-\Delta^2}}
\theta(\delta\mu-\Delta) \nonumber \\
&\simeq & \frac{4\alpha_s \bar\mu^2 \delta\mu}
{\pi\sqrt{(\delta\mu)^2-\Delta^2}}\theta(\delta\mu-\Delta),
\label{m_D_1}
\end{eqnarray}
where $\alpha_s\equiv g^2/4\pi$, $\bar\mu=(\mu_{ur}+\mu_{dg})/2$, 
and $\mu^{\pm}=\bar\mu\pm\sqrt{(\delta\mu)^2-\Delta^2}$ are the
values of the ``effective'' Fermi momenta of the gapless quasiparticles
\cite{SH,HS}. The mass as a function of $\Delta/\delta\mu$ is 
shown in the upper panel of Fig.~\ref{fig-Masses} with a red solid 
line. It is easy to check that the result in Eq.~(\ref{m_D_1}) 
reduces to the known expressions in the normal phase 
(i.e., $\Delta/\delta\mu=0$) \cite{Vija,Manuel} and in the ideal 
2SC phase (i.e., $\Delta/\delta\mu=\infty$) \cite{R-meissner}.

We note that the Debye screening mass in Eq.~(\ref{m_D_1}) vanishes 
in the gapped phase (i.e., $\Delta/\delta\mu>1$). As in the case of 
the ideal 2SC phase, this reflects the fact that there are no gapless 
quasiparticles charged with respect to the unbroken SU(2)$_c$ gauge 
group. In the gapless 2SC phase, such quasiparticles exist and the 
value of the Debye screening mass is proportional to the density of
states at the corresponding ``effective'' Fermi surfaces 
\cite{HS,kemer}. 

In the limit when the ratio $\Delta/\delta\mu$ approaches $1$ from
below (i.e., from the  gapless phase), the formal value of the 
Debye mass in Eq.~(\ref{m_D_1}) goes to infinity. This 
is because the density of the gapless states goes to infinity at 
$\Delta/\delta\mu=1$ when the gapless quasiparticle dispersion 
relation becomes quadratic, i.e., $E_{\Delta^{-}}^{-}\simeq (p-\bar\mu)^2
/2\Delta$ in notation of Ref.~\cite{HS}.

The Meissner screening mass is defined by Eq.~(\ref{def-Meissner}).
An explicit calculation for the gluons of the unbroken SU(2)$_c$ 
shows that their Meissner masses are vanishing in the gapped and 
gapless 2SC phases,
\begin{equation}
m_{M,1}^2 =0 . 
\label{m_M_1}
\end{equation}
Of course, this is in agreement with the general group-theoretical 
arguments.

{\em Screening masses of the gluons with $A=4,5,6,7$.}
The polarization tensor $\Pi_{AB}^{\mu\nu}(0,p)$ with $A,B=4,5,6,7$ 
is not diagonal. It becomes diagonal after switching to the physical 
basis in terms of $A_{4\pm}^{\mu}=(A_{4}^{\mu}\pm i A_{5}^{\mu})/\sqrt{2}$
and $A_{6\pm}^{\mu}=(A_{6}^{\mu}\pm i A_{7}^{\mu})/\sqrt{2}$ fields. 
In the static limit, all four eigenvalues of the polarization tensor 
are degenerate. By making use of the definition in Eq.~(\ref{def-Debye}), 
we derive the following result for the corresponding Debye masses:
\begin{equation}
m_{D,4}^2 = \frac{4\alpha_s\bar\mu^2}{\pi} \left[
\frac{\Delta^2+2\delta\mu^2}{2\Delta^2} 
-\frac{\delta\mu\sqrt{\delta\mu^2-\Delta^2}}{\Delta^2}
\theta(\delta\mu-\Delta)\right] \!\! .
\label{m_D_4}
\end{equation}
Here we assumed that $\mu_8$ is vanishing which is a good 
approximation in neutral two-flavor quark matter \cite{HS,DD}.
The complete result with nonzero $\mu_8$ will be reported 
elsewhere \cite{pi-long}.

Now, the fourfold degenerate Meissner screening mass of the gluons 
with $A=4,5,6,7$ reads
\begin{equation}
m_{M,4}^2 = \frac{4\alpha_s\bar\mu^2}{3\pi} \left[
\frac{\Delta^2-2\delta\mu^2}{2\Delta^2} 
+\frac{\delta\mu\sqrt{\delta\mu^2-\Delta^2}}{\Delta^2}
\theta(\delta\mu-\Delta)\right] \!\! .
\label{m_M_4}
\end{equation}
Both results in Eqs.~(\ref{m_D_4}) and (\ref{m_M_4}) interpolate between 
the known results in the normal phase (i.e., $\Delta/\delta\mu=0$) and 
in the ideal 2SC phase (i.e., $\Delta/\delta\mu=\infty$) \cite{R-meissner}. 
The screening masses are plotted in Fig.~\ref{fig-Masses} using green 
long-dashed lines.

The most interesting observation about the gluon screening properties 
here is that the expression on the right hand side of Eq.~(\ref{m_M_4}) 
is {\em negative} when $0<\Delta/\delta\mu<\sqrt{2}$. The standard 
interpretation of such a result is the existence of a plasma type 
instability in the system \cite{Mrow1,RS,Arnold,Mrow2}. In the case
at hand, we find that the instability appears in the whole family of 
gapless 2SC phases (with $0<\Delta/\delta\mu<1$) and even in some 
gapped 2SC phases (with $1<\Delta/\delta\mu<\sqrt{2}$). 

While one may argue that the plasma instability in the gapless phase 
is related to the instability discussed in Ref.~\cite{WuYip}, it 
is obviously not the case in the gapped 2SC phase when
$1<\Delta/\delta\mu<\sqrt{2}$. The arguments of Ref.~\cite{WuYip} 
are based on the fact that the ground state corresponds to a local
maximum of the effective potential in the system with a fixed 
mismatch parameter, $\delta\mu=\mbox{const}$. However, without 
imposing the charge neutrality ($\delta\mu=\mbox{const}$), the 
gapped 2SC ground state with $\Delta/\delta\mu \in (1,\sqrt{2})$ 
corresponds to a local minimum of the effective potential. Although 
it is not the global minimum for $\Delta/\delta\mu \in (1,\sqrt{2})$,
it is unlikely that the gluon screening masses could probe 
the global structure of the effective potential. Thus, the true 
origin of the gluon instability is yet to be clarified. 

{\em Screening masses and mixing of the 8th gluon and the photon.}
From the general arguments, one knows that there is a new 
$\tilde{\rm U}(1)_{\rm em}$ symmetry in the 2SC/g2SC phase. 
The new medium photon is a mixture of the 8th gluon and the 
vacuum photon (the corresponding generator is $\tilde{Q} = Q - 
\frac{1}{\sqrt{3}}  T_8$).

It is important to emphasize that the explicit result for the
$00$-components of the polarization tensor of the 8th gluon and 
the {\em vacuum} photon have no mixing at $p_0=0$ and $p\to 0$. 
This is similar to the ideal 2SC case considered in Ref.~\cite{SWR}. 
(In view of this, one should be careful when interpreting the 
results for the Debye screening masses in a different basis 
of gauge fields \cite{Litim}.) The expressions for the Debye 
screening masses read
\begin{eqnarray}
m_{D,8}^2 &=& \frac{4\alpha_s\bar\mu^2}{\pi} ,
\label{m_D_8} \\
m_{D,\gamma}^2 &=& \frac{8\alpha\bar\mu^2}{3\pi} 
\left(1+\frac{3\delta\mu~\theta(\delta\mu-\Delta)}
{2\sqrt{(\delta\mu)^2-\Delta^2}} \right),
\label{m_D_gamma} 
\end{eqnarray}
where $\alpha\equiv e^2/4\pi$ is the fine structure constant.

In order to obtain the Meissner screening masses, we first derive
all the components of the polarization tensor that span the space  
of the 8th gluon and the {\em vacuum} photon. At $p_0=0$ and 
$p\to 0$, the corresponding nonzero components, denoted as 
$m_{M,AB}^2$ in analogy with Eq.~(\ref{def-Meissner}), read
\begin{eqnarray}
m_{M,88}^2 &=& \frac{4\alpha_s\bar\mu^2}{9\pi} 
\left(1-\frac{\delta\mu~\theta(\delta\mu-\Delta)}
{\sqrt{(\delta\mu)^2-\Delta^2}} \right),
\label{m_M_88} \\
m_{M,\gamma\gamma}^2 &=& \frac{4\alpha\bar\mu^2}{27\pi} 
\left(1-\frac{\delta\mu~\theta(\delta\mu-\Delta)}
{\sqrt{(\delta\mu)^2-\Delta^2}} \right) ,
\label{m_M_8-gamma} \\
m_{M,8\gamma}^2 &=& 
\frac{4\sqrt{\alpha\alpha_s}\bar\mu^2}{9\sqrt{3}\pi} 
\left(1-\frac{\delta\mu~\theta(\delta\mu-\Delta)}
{\sqrt{(\delta\mu)^2-\Delta^2}} \right),
\label{m_M_gamma-gamma}
\end{eqnarray}
and $m_{M,\gamma 8}^2 = m_{M,8 \gamma}^2 $. The mixing disappears 
in the basis of the fields,
\begin{eqnarray}
\tilde{A}^8_\mu = A^8_\mu \cos\varphi + A^\gamma_\mu \sin\varphi ,\\
\tilde{A}^\gamma_\mu = A^\gamma_\mu \cos\varphi  - A^8_\mu \sin\varphi ,
\end{eqnarray}
where the mixing angle $\varphi$ is determined 
by the symmetry arguments,
\begin{equation}
\sin\varphi = \sqrt{\frac{\alpha}{3\alpha_s+\alpha}}.
\end{equation}
As one might have expected, the mixing angle is the same in the 
gapless and the gapped phases. Of course, it also coincides with 
the mixing angle in the ideal 2SC case in Ref.~\cite{SWR}.

It should be noted that the absence of a mixing between the 
electrical modes of the 8th gluon and the vacuum photon in the
static limit, see Eqs.~(\ref{m_D_8}) and (\ref{m_D_gamma}), 
is not in conflict with the gauge invariance of the 2SC/g2SC 
ground state with respect to $\tilde{\rm U}(1)_{\rm em}$.
The two propagating modes of the low-energy photon $\tilde{\gamma}$ 
of $\tilde{\rm U}(1)_{\rm em}$ should have transverse polarizations
and, therefore, should come from the magnetic sector. The third, 
electrical mode of $\tilde{\gamma}$ is not massless. It decouples
from the low-energy theory and its presence is irrelevant for the 
gauge invariance with respect to $\tilde{\rm U}(1)_{\rm em}$.

The Meissner screening mass for the new gluon field is
\begin{equation}
m_{M,\tilde{8}}^2 = \frac{4(3\alpha_s+\alpha)\bar\mu^2}{27\pi} 
\left(1-\frac{\delta\mu~\theta(\delta\mu-\Delta)}
{\sqrt{(\delta\mu)^2-\Delta^2}} \right),
\label{m_M_8} 
\end{equation}
which is shown in the lower panel of Fig.~\ref{fig-Masses} using a 
blue short-dashed line.
The Meissner screening mass for the new photon $\tilde{\gamma}$ 
is vanishing, $m_{M,\tilde{\gamma}}^2 = 0$. This is consistent with 
the absence of the Meissner effect for the 
unbroken $\tilde{\rm U}(1)_{\rm em}$.

As is easy to see from Eq.~(\ref{m_M_8}), the medium modified
$\tilde{8}$th gluon has a magnetic plasma instability in the gapless 
2SC phase. This is because the Meissner screening mass squared is 
{\em negative} when $0<\Delta/\delta\mu<1$.

\section{Discussion}
\label{discussion}

In this paper, we calculated the Debye and Meissner screening masses 
of the gluons and the photon in the case of neutral, $\beta$-equilibrated 
dense two-flavor quark matter (see Fig.~\ref{fig-Masses}). Our 
results interpolate between the 
known values in the normal phase \cite{Vija,Manuel} and in the ideal 
2SC phase \cite{R-meissner,SWR} of quark matter. 

The Meissner screening properties of dense matter are most interesting. 
We find that there is a chromomagnetic plasma type instability in quark 
matter \cite{footnote1}. 
This is driven by 5 unstable gluon modes ($A=4,5,6,7,\tilde{8}$) 
in the g2SC phase ($0<\Delta/\delta\mu<1$), and by 4 modes ($A=4,5,6,7$) 
in the gapped 2SC phase when $1<\Delta/\delta\mu<\sqrt{2}$. One could 
connect the instability from the $\tilde{8}$th gluon to a large density 
of gapless states in the g2SC phase. This might be related to the 
mechanism proposed in Ref.~\cite{WuYip}. The other 4 unstable modes 
have no obvious connection with the existence of the g2SC phase 
because they lead to an instability even in the gapped phase. 

It is natural to assume that the instability, indicated by the negative 
values of the Meissner screening masses squared, will resolve in some 
type of a gluon condensation. Indeed, the result $m_{M,A}^2<0$ suggests 
that the system stays in a false vacuum that corresponds to a local 
maximum of the effective potential for the gluon field. Then, the true 
vacuum is most likely given by the global minimum of the potential that 
corresponds to a nonzero expectation value of a gluon field, i.e., 
$\langle A_\mu^A \rangle\neq 0$ for $A\in (4,5,6,7)$ or for $A=\tilde{8}$, 
depending on which instability develops first. 

We cannot exclude the possibility that the true ground state 
has a condensate that breaks the rotational symmetry of the system. 
In fact, this would be the most natural outcome of a gluon 
condensation because it is the magnetic components of gluons 
$A_i^A$ that drive the instability. One might even speculate 
\cite{footnote2} that the mechanism is similar to that in 
Ref.~\cite{GMS}. 

In passing, we note that the gluon type instability, indicated by 
negative values of the Meissner screening masses squared, is
not directly related to the so-called Sarma instability \cite{Sarma}. 
As was shown in Ref.~\cite{SH}, the Sarma instability in the 
effective potential for the order parameter is removed when the 
neutrality condition is imposed. 

The nature of the instability observed here may 
resemble the instability in anisotropic models of 
Refs.~\cite{Mrow1,RS,Arnold,Mrow2} used for describing the
initial stage of heavy ion collisions. However, the origins of
the two instabilities are very different. In contrast to the situation in 
Refs.~\cite{Mrow1,RS,Arnold,Mrow2}, the quark distribution functions 
are completely isotropic in momentum space in neutral dense quark matter.

In the future, it would be very interesting to investigate whether 
a similar instability develops in the gCFL phase \cite{gCFL,RSR}, 
where the low-energy quasiparticle spectrum resembles the spectrum 
in the g2SC phase.

{\bf Acknowledgments.}
The authors thank M.~Buballa, M.~Forbes, T.~Hatsuda, D.~Hou, A.~Iwazaki, 
T.~Koide, C.~Kouvaris, J.~Lenaghan, V.~Miransky, A.~M\'ocsy, 
S.~Mr\'owczy\'nski, R.~Pisarski, K.~Rajagopal, A.~Rebhan, 
P.~Reuter, D.~Rischke, P.~Romatschke, T.~Sch\"afer, A.~Schmitt, 
D.~Son, M.~Strickland, M.~Tachibana, D.~N.~Voskresensky, and 
Q.~Wang for interesting 
discussions. I.A.S. is grateful to the INT at the University of 
Washington in Seattle for its hospitality. The work of M.H. was 
supported by the Alexander von Humboldt-Foundation, and by the 
NSFC under Grants No. 10105005 and No. 10135030. The work of 
I.A.S. was supported by Gesellschaft f\"{u}r Schwerionenforschung 
(GSI) and by Bundesministerium f\"{u}r Bildung und Forschung (BMBF).


\begin{thebibliography}{99}
\bibitem{footnote1} Note that the plasma instability may not 
develop if the gluons are massive because of non-perturbative 
effects.
\end{thebibliography}
\end{document}